\documentclass[preprint,prl,showpacs,byrevtex,superscriptaddress]{revtex4-1}
\usepackage{graphicx}
\usepackage{epsfig}

\usepackage[caption=false]{subfig}
\usepackage{amsmath,braket,relsize}
\usepackage{amssymb}
\usepackage{bm}
\usepackage{color}

\begin{document}

\title{Hybrid Two-Qubit Gate using Circuit QED System with Triple-Leg Stripline Resonator}
\author{Dongmin Kim}
\author{Kyungsun Moon$^*$}
\affiliation{Department of Physics and IPAP, Yonsei University, Seoul 03722, Korea, *E-mail: kmoon@yonsei.ac.kr}

\begin{abstract}

We theoretically propose a circuit QED system implemented with triple-leg stripline resonator (TSR). Unlikely from linear stripline resonator, the fundamental intra-cavity microwave modes of the TSR are two-fold degenerate. When a superconducting qubit is placed near one of the TSR legs, one fundamental mode is directly coupled to the qubit, while the other one remains uncoupled. Our system closely resembles an optical cavity QED system, where an atom in a cavity couples only to the incident photon with a specific polarization by placing a polarization beamsplitter in front of the optical cavity.

Using our circuit QED system, we have theoretically studied a two-qubit quantum gate operation in a hybrid qubit composed of flying microwave qubit and superconducting qubit. We have demonstrated that for the hybrid qubit, the quantum controlled phase flip (CPF) gate can be reliably implemented for the experimentally available set of parameters.

\end{abstract}

\maketitle

Circuit QED systems consist of a superconducting qubit located inside a high-Q superconducting linear stripline resonator (LSR) or a three-dimensional waveguide cavity, which have experimentally demonstrated a strong coupling between superconducting qubit and intra-cavity microwave photon. This has made possible the coherent control of a superconducting qubit by a single photon and \textit{vice versa}, which has been successfully applied to quantum computing and quantum optics\cite{Blais, Transmon, Transmon2, Moon, Rhen, Abdumalikov, Manucharyan, Paik, Mi}.
As a representative of flying qubits, optical qubits are the most advanced qubits to realize the future quantum technologies such as optical quantum computation, long-distance quantum communication, and quantum cryptography\cite{Reiserer, Resierer2, Shomroni, KLM}.
They can be encoded with the superposition of any two degrees-of-freedom of a photon: Polarization, path, time-bin, and frequency-bin\cite{Hofmann,Nemoto,Kwon,Zhang,Ramelow}. Coherent superposition of two optical coherent states with equal absolute amplitude but opposite sign can also be used as a qubit\cite{Ralph,Lund}.
Much effort has been taken to realize two-photon quantum logic gate\cite{Koshino,Gorshkov,Paredes-Barato,Khazali,Hacker}.
However, photon-photon interaction is not strong enough to implement a reliable two-photon quantum logic gate.
Using the optical cavity QED system, there have been several interesting studies to implement two-qubit quantum gates for Rydberg atoms\cite{Saffman,Muller,Beterov,Das}.

\begin{figure}
\epsfxsize=6in \epsfysize=3.5in \epsffile{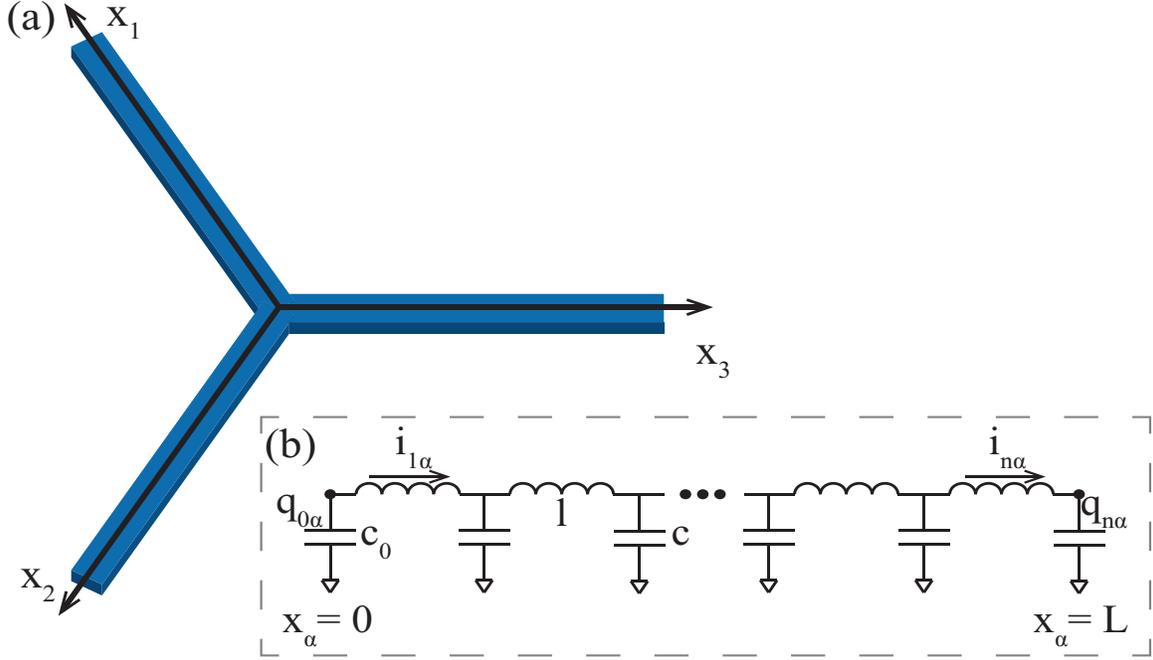}
\caption{(a) Schematic diagram of the triple-leg stripline resonator system. Three linear striplines with identical length are joined at the center. (b) Lumped element circuit model for the TSR. Capacitance of the central capacitor is taken to be $c_0$. In the circuit, $i_{m\alpha}$ represents the current flowing through the $m$-th inductor with inductance $l$ and $q_{m\alpha}$ represents the charge accumulated on the $m$-th capacitor with capacitance $c$ in the $\alpha$-th leg.}
\label{TSR1}
\end{figure}

In the paper, we theoretically propose a new circuit QED system, where a superconducting qubit is coupled to the intra-cavity microwave photons inside the TSR. The schematic diagram of the TSR with three legs are illustrated in Fig. \ref{TSR1}(a). Circuit QED systems are generally implemented with the LSR, where the intra-cavity microwave modes are non-degenerate. In contrast, the fundamental intra-cavity microwave modes of the TSR are two-fold degenerate with distinct spatial mode structure. When a superconducting qubit is placed at one of the three legs, the degeneracy is lifted. We have shown that one mode is coupled to the qubit, while the other mode remains uncoupled.
Our system closely resembles a special optical cavity QED system, where an atom in a cavity couples only to the incident photon with a specific polarization.
In order to achieve a polarization-selective coupling to the incident photon, they have placed a polarization beamsplitter in front of the optical cavity\cite{Duan}.
Using our circuit QED system, we want to implement the hybrid two-qubit gate between flying microwave qubit and superconducting qubit.
While the photon in a flying optical qubit can have a definite polarization, the intra-cavity microwave photon inside the LSR/TSR does not have a polarization mode but a distinct spatial mode structure as a standing electromagnetic wave.
In the work, we propose the flying microwave qubit to be a coherent superposition of two individual coherent states or a dual-rail train of photon pulses synchronized in time, but spatially separated in the transverse direction. Individual coherent state or single photon pulse has a spatio-temporal mode structure with identical profile and center frequency, which is comparable to the fundamental frequency of the intra-cavity microwave photon.
By setting up the special optical circuit, we have been able to couple the flying microwave qubit to the two fundamental intra-cavity microwave modes. This in turn makes possible to realize the hybrid two-qubit gate composed of flying microwave qubit and superconducting qubit.
Using the standard input-output theory, we have theoretically demonstrated that the quantum CPF gate for the hybrid qubit can be reliably implemented for the experimentally available set of parameters.

\begin{figure}
\epsfxsize=3.5in \epsfysize=3.5in \epsffile{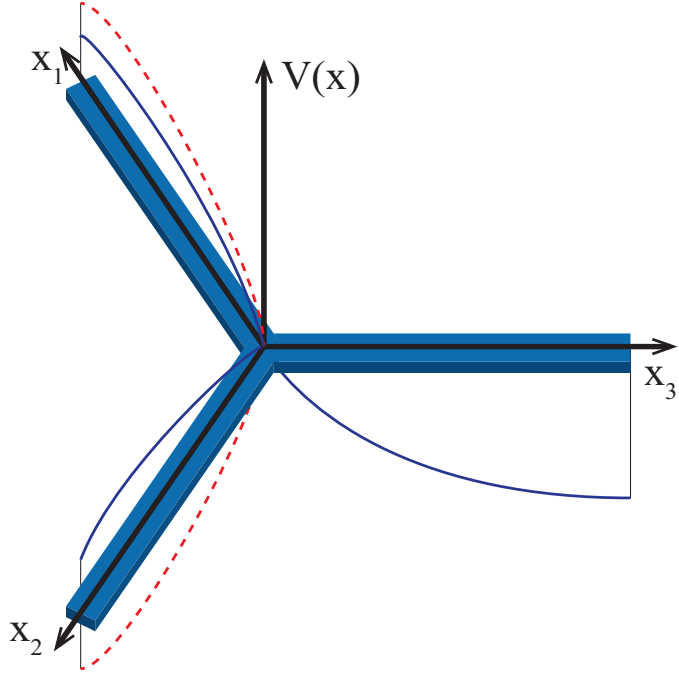}
\caption{Spatial mode structure of the two-fold degenerate intra-cavity microwave modes for voltage $V(x)$ in the TSR system. The blue solid curve represents the $(+)$-mode with eigenvector ${\bf A}^+=(1/\sqrt{6},1/\sqrt{6},-2/\sqrt{6})$ and the red dashed curve represents the $(-)$-mode with eigenvector ${\bf A}^-=(1/\sqrt{2},-1/\sqrt{2},0)$. A single node is located at the center and the three antinodes are located at the ends of each leg. For the $(-)$-mode, the voltage is zero at the third leg.}
\label{TSR2}
\end{figure}

We will describe our TSR system using the lumped element circuit model as shown in Fig. \ref{TSR1}(b), whose Lagrangian can be written by
\begin{equation}	
{\cal L}\left[\left\{q_{n\alpha}\right\}\right]=\sum_{\alpha=1}^3\sum_{n=1}^N\left[{l\over 2} i_{n\alpha}^2-{1\over 2c} q_{n\alpha}^2\right]-{1\over 2c_0} q_0^2
\end{equation}
where $q_{n\alpha}$ represents the charge accumulated on the capacitor at the position $n$ of the $\alpha$-th leg with $\alpha=1,2,3$, $i_{n\alpha}$ the current flowing through the inductor at the position $n$ of the $\alpha$-th leg, $q_0$ the charge on the central capacitor with capacitance $c_0$\cite{Blais, Rhen}.
By introducing the non-local variables $\theta_{m\alpha}=\sum_{n=m}^N q_{n\alpha}$, one can automatically satisfy the Kirchhoff rules, which the local currents and charges should obey. Using the variables $\theta_{m\alpha}$, the Lagrangian of the system can be written by
\begin{equation}	
{\cal L}\left[\left\{\theta_{m\alpha}\right\}\right]=\sum_{\alpha=1}^3\sum_{n=1}^N\left[{l\over 2} {\dot\theta}_{n\alpha}^2-{1\over 2c} \left(\theta_{n\alpha}-\theta_{n+1\alpha}\right)^2\right] -{1\over 2c_0}\left(\sum_{\alpha=1}^3\theta_{1\alpha}\right)^2
\label{DiscreteLag}
\end{equation}
where we have imposed the charge neutrality condition.
For $n>1$, the equation of motion for the field $\theta_{n\alpha}(t)$ will follow the wave equation with the boundary condition of $\theta_{N+1\alpha}(t)=0$.
In the continuum limit, the spatial part of the continuum field $\theta_{\alpha}(x,t)=a_0^{-1}\int_x^L dx^\prime q_\alpha(x^\prime,t)$ can be given by $\chi_\alpha(x)=A_\alpha \sin k(x-L)$ with $k=\omega/v_p$ and $v_p=a_0/\sqrt{lc}$. Here the $\alpha$-th leg extends from the center of the TSR at $x_\alpha=0$ to $x_\alpha=L$ and $L, a_0$ denote the total length of each leg and the unit length, respectively.
The equation of motion for the field $\theta_{1\alpha}(t)$ will lead to the following differential equations for $\chi_\alpha(0)$
\begin{equation}	
{a_0\over c}{\partial\chi_\alpha(0)\over \partial x}-{1\over c_0}\sum_{\beta=1}^3 \chi_\beta(0)+l\omega^2\chi_\alpha(0)=\left({a_0\over c}k\cos kL-l\omega^2\sin kL\right)A_\alpha+{1\over c_0}\sin kL \sum_{\beta=1}^3 A_\beta=0.
\end{equation}
For $\sum_{\beta=1}^3 A_\beta\ne 0$, $\tan kL = ka_0/[(ka_0)^2-3c/c_0]\cong -(c_o/3c)ka_0$. For $k\cong (n+1/2)\pi/L$, $\tan kL = 1/(ka_0)$.
For the low-frequency limit ($ka_0<<1$), there exists a symmetric mode with $k\cong n\pi/L$ for $n$ being a natural number and $A_\alpha=1/\sqrt{3}$ for every $\alpha$. In addition, there exist two-fold degenerate modes with $k\cong (n-1/2)\pi/L$ and $\sum_{\alpha=1}^3 A_\alpha=0$.
For the two-fold degenerate modes, one can choose any two linearly independent basis, which satisfy the condition $\sum_{\beta=1}^3 A_\beta=0$. For example, one can choose
${\bf A}^+=(1/\sqrt{6},1/\sqrt{6},-2/\sqrt{6})$ and ${\bf A}^-=(1/\sqrt{2},-1/\sqrt{2},0)$ for two orthonormal eigenvectors as shown in Fig \ref{TSR2}. For the two-fold degenerate fundamental harmonic modes ($n=1$), the two normalized spatial basis functions can be written by $\phi_\pm(x)=\sqrt{2/3L}\sum_{\alpha=1}^3 A_\alpha^\pm\sin \left[\pi(x_\alpha-L)/2L\right]$.
By quantizing the classical Hamiltonian of the TSR system, one can obtain the following quantum Hamiltonian describing the lowest-energy fundamental harmonic modes
\begin{equation}
H=\hbar\omega_0\left({\hat a}^\dag_{+} {\hat a}_{+}+{\hat a}^\dag_{-} {\hat a}_{-}\right)
\end{equation}
where $\omega_0=v_p(\pi/2L)$.
Unlikely from the LSR, our TSR system can indeed support the fundamental intra-cavity modes, which are two-fold degenerate.
Here the direct two-photon interactions are presumed to be quite weak.

We now place a superconducting qubit near the position $x$ of the TSR, which induces a capacitive coupling between the superconducting qubit and the TSR.
We want to study the coupled system of a superconducting qubit and the intra-cavity microwave photons. In terms of creation and annihilation operators ${\hat a}_\pm^\dag, {\hat a}_\pm$, the quantized field ${\hat\theta}(x)$ can be written by
\begin{equation}
{\hat\theta}(x)=\sum_{\mu=\pm}\sum_{\alpha=1}^3 {A_\alpha^\mu \sqrt{\frac{\hbar a_0}{3l\omega_0 L}}\sin{k_0 (x_\alpha -L)}({\hat a}_\mu+{\hat a}_\mu^\dag)}
\end{equation}
where $x$ will denote the position $x_\alpha$, when the superconducting qubit is placed near position $x_\alpha$ of the $\alpha$-th leg, and $k_0=\pi/2L$. The quantized voltage ${\hat V}(x)$ can be obtained by applying the following formula
\begin{equation}
{\hat V}(x)=\frac{{\hat q}(x)}{c}=-\frac{a_0}{c}\frac{\partial{\hat\theta}}{\partial x}=\sum_{\mu=\pm}\sum_{\alpha=1}^3 g_\mu (x_\alpha) \left({\hat a}_\mu+{\hat a}_\mu^\dag\right)
\end{equation}
where the function $g_\mu(x_\alpha)$ represents the coupling strength between superconducting qubit and intra-cavity microwave photon with mode $\mu$, which is given by
\begin{equation}
g_\mu (x_\alpha) = -A_\alpha^{\mu} \sqrt{\frac{\hbar \omega_0 a_0}{3Lc}}\cos k_0(x_\alpha -L).
\end{equation}
The coupling strength $g_\mu(x_\alpha)$ can be adjusted by varying the position of the superconducting qubit.
We will assume that the superconducting qubit is placed near the position $x_3$ of the third leg. In this case, $g_+=\sqrt{2\hbar\omega_0 a_0/9Lc}\cos k_0(x_3 -L)$ and $g_-=0$.

Through the capacitative coupling between a superconducting qubit and two-degenerate intra-cavity microwave photons, the Hamiltonian for a coupled system can be written by
\begin{equation}
{\hat H}=\hbar\omega_0({\hat a}^\dag_{+} {\hat a}_{+}+{\hat a}^\dag_{-} {\hat a}_{-}) +\frac{E}{2}\sigma_z
+g_+(\sigma^+{\hat a}_{+}+\sigma^-{\hat a}^\dag_{+}).
\end{equation}
One can clearly notice that only the $(+)$-mode couples to the superconducting qubit, while the $(-)$-mode does not couple. The preferential coupling of a superconducting qubit to the $(+)$-mode alone is quite similar to the case of an atomic system, where only right/left circularly polarized photon can interact with the atom due to total angular momentum conservation\cite{KLM}. Our system also closely resembles an optical cavity QED system, where an atom in a cavity couples only to the horizontal/vertical polarized photon mode alone\cite{Duan}.
Hence the two-fold degeneracy of the fundamental intra-cavity microwave modes is lifted in the presence of a capacitive coupling to the superconducting qubit.

Using the unitary transformation ${\hat U}=\exp{[(g/\Delta)({\hat a}_+^\dag\sigma^- -{\hat a}_+\sigma^+)]}$, one can obtain the following effective Hamiltonian in the dispersive regime of $g/\vert\Delta\vert\ll 1$
\begin{equation}
{\hat H}_{\rm eff}={\hat U}^\dag {\hat H} {\hat U}=\hbar\omega_{0}\left({\hat a}^\dag_+ {\hat a}_+ +{\hat a}^\dag_- {\hat a}_-\right) +\frac{E}{2}\sigma_z+\chi {\hat a}^\dag_+{\hat a}_+\sigma_z
\label{eq:Hamiltonian}
\end{equation}
where $\Delta=E-\hbar\omega_0$ represents the detuning between the frequency of the superconducting qubit and that of the fundamental intra-cavity microwave photon, and $\chi=g^2/\Delta$.
Hence the frequency of the $(+)$-mode is blue- or red-detuned by $\chi\sigma_z/\hbar$ depending on the qubit state of $\sigma_z=\pm 1$,
while that of the $(-)$-mode is unchanged.
We will utilize this interesting property of our circuit QED system to the quantum information processing. Especially we want to implement a hybrid two-qubit gate between flying microwave qubit and superconducting qubit.

\begin{figure}
\epsfxsize=6in \epsfysize=3.5in \epsffile{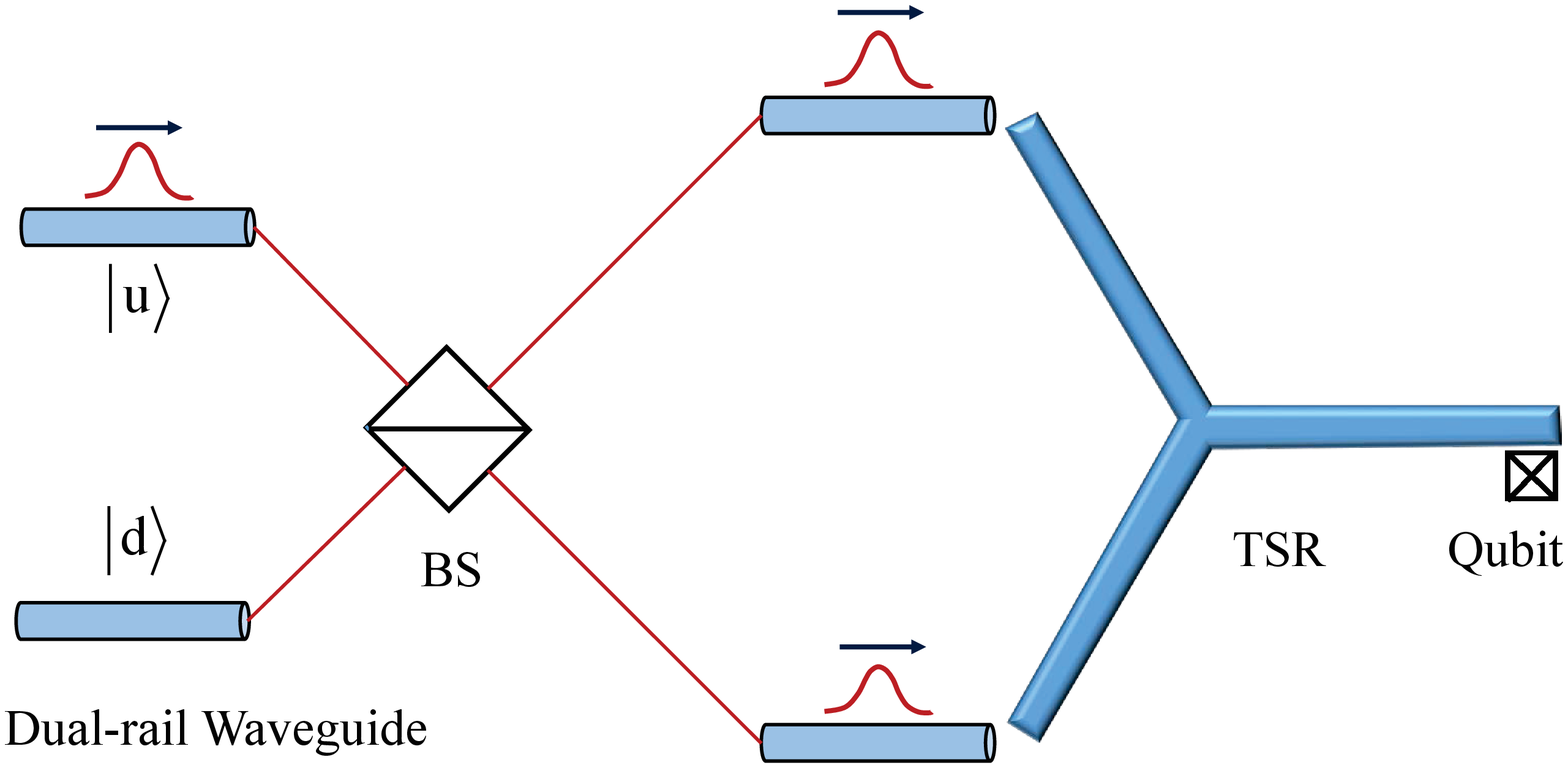}
\caption{An optical circuit to couple the flying microwave qubit to the two intra-cavity microwave photon modes. A superconducting qubit is placed at the end of the third leg. Input photon pulses from the dual-rail waveguide are split into two components by the beamsplitter(BS), which are fed into the leg 1 and leg 2 of the TSR, respectively.}
\label{TSR3}
\end{figure}

In Fig. \ref{TSR3}, we have proposed an optical circuit, which is composed of a dual-rail waveguide, a beamsplitter, and the circuit QED system with the TSR. We propose that the flying microwave qubits are a coherent superposition of two individual coherent states or dual-rail train of photon pulses, which are synchronized in time, but spatially separated in the transverse direction by the dual-rail waveguide.
We assume that individual coherent state or photon pulse in a given flying microwave qubit has a spatio-temporal mode structure with identical profile and center frequency. The general quantum state of a flying microwave qubit can be described by $\ket{\eta,\phi} = \sqrt{\eta}\ket{\rm u} + e^{i\phi}\sqrt{1-\eta} \ket{\rm d}$, where $\ket{\rm u}(\ket{\rm d})$ represents the individual quantum state spatially localized at the upper(lower)-rail. One can perform arbitrary single qubit rotations by using a beamsplitter and phase shifter\cite{Ralph}. The flying optical qubits encode the quantum information as a time-sequence of input pulses, where each pulse represents either one of $\ket{\rm u}$ and $\ket{\rm d}$.
One can define the creation operators for the diagonal-basis states $\ket{\rm u}$ and $\ket{\rm d}$ as follows: ${\hat b}_{\rm u}^\dagger\ket{0}=\ket{\rm u}, {\hat b}_{\rm d}^\dagger\ket{0}=\ket{\rm d}$.
In order to couple the flying microwave qubit to the intra-cavity microwave photons, we have placed a 50/50 beamsplitter as demonstrated in Fig. \ref{TSR3}. The 50/50 beamsplitter splits the input pulses, which can be described by the following operator transformations
\begin{equation}
{\hat b}_{\rm u}^+={1\over \sqrt{2}}\left({\hat b}_1^\dag+{\hat b}_2^\dag\right), \,\,\,\,\,\,{\hat b}_{\rm d}^+={1\over \sqrt{2}}\left({\hat b}_1^\dag-{\hat b}_2^\dag\right)
\end{equation}
where the minus sign in the second equation comes from the reflection phase shift of $\pi$ and ${\hat b}_1^\dag, {\hat b}_2^\dag$ represent the creation operators for the input microwave photons fed into the leg 1 and leg 2 of the TSR, respectively.
The total Hamiltonian of the circuit QED system plus the external input pulses can be written by
\begin{equation}
{\hat H}_{\rm tot}={\hat H}_{\rm eff}+\sum_{i=1,2}\omega_r {\hat b}^\dag_i {\hat b}_i+\sum_{i=1,2}\lambda_i\left({\hat a}^\dag_i {\hat b}_i+{\hat a}_i {\hat b}^\dag_i\right)
\end{equation}
where ${\hat a}^\dag_i$ represents the creation operator for the intra-cavity microwave photon at the $i$-th leg and $\lambda_i$ represents the coupling strength between the input microwave photon and the intra-cavity microwave photon at the $i$-th leg.
We will assume that the optical elements used in our optical circuit are lossless, which is a reasonable assumption for modern optical components.

Following the discrete rotational symmetry of the TSR, we have also assumed that the coupling strengths to the leg 1 and leg 2 can be taken to be identical in principle such that $\lambda_1=\lambda_2=\lambda$ and then the total Hamiltonian can be written by
\begin{equation}
{\hat H}_{\rm tot}={\hat H}_{\rm eff}+\sum_{\alpha={\rm u,d}}\omega_r {\hat b}^\dag_\alpha {\hat b}_\alpha+\lambda\left({\hat a}^\dag_+ {\hat b}_{\rm u}+{\hat a}_-^\dag {\hat b}_{\rm d} + {\rm h.c.}\right).
\end{equation}
Hence the incident flying microwave photon in the state $\ket{\rm u}(\ket{\rm d})$ is directly coupled to the intra-cavity microwave photon with the $(+)((-))$ mode.

One currently active field of research is the use of linear optics versus the use of nonlinear optics in quantum computing. The KLM protocol is an implementation of linear optical quantum computing (LOQC), which makes it possible to create universal quantum computers only with linear optical tools\cite{KLM}. The KLM protocol uses linear optical elements, single photon sources, and photon detectors as resources to construct a quantum computation scheme involving only ancilla resources, quantum teleportations, and error corrections. As an elemental gate in the KLM protocol, one needs to implement the two-photon quantum CPF gate.
It converts the initial two-photon state $\ket{\psi_i}=\left(\ket{\rm uu}+\ket{\rm ud}+\ket{\rm dd}+\ket{\rm du}\right)/2$ to the final two-photon state $\ket{\psi_f}=\left(\ket{\rm uu}+\ket{\rm ud}+\ket{\rm dd}-\ket{\rm du}\right)/2$.
Here we will propose a method to realize the two-photon quantum CPF gate using our circuit QED system. The initial states of the flying microwave qubits are prepared as time-sequences of gaussian photon pulses encoded with the definite mode indices of u and d and the duration of individual pulse is given by $T$.
We will apply the input-output theory, where the input beam is taken to be the gaussian photon pulses as described above.
In the Heisenberg picture, the equation of motion for a given quantum observable ${\hat{\cal O}}(t)$ in the Lindblad form can be written by
\begin{equation}
{\frac{d{\hat{\cal O}}}{dt}}={\frac {i}{\hbar }}[{\hat H}_{\rm eff},{{\hat{\cal O}}}]+\sum _{i=1}^3\gamma_i\left({\hat L}_{i}^{\dagger }{{\hat{\cal O}}}{\hat L}_{i}-{\frac {1}{2}}\left\{{\hat L}_{i}^{\dagger }{\hat L}_{i},{{\hat{\cal O}}}\right\}\right)
\end{equation}
where ${\hat L}_1=\sigma^-$ with $\gamma_1=\gamma$ representing a spontaneous emission rate, ${\hat L}_2=\sigma_z$ with $\gamma_2=\gamma_\varphi/2$ and $\gamma_\varphi$ representing a part of the dephasing rate, and ${\hat L}_3={\hat a}$ with $\gamma_3=\kappa$ representing the cavity loss\cite{Clerk,Walls}. The operators ${\hat a}(t)$, $\sigma_z(t)$, and $\sigma_+(t)$ will satisfy the following set of equations with the input photon pulse described by the operator ${\hat a}_{\rm in}(t)$
\begin{eqnarray}
{\frac{da_+}{dt}}&=&-i\left(\omega_0+\chi\sigma_z(t)\right){\hat a}_+(t)-{\frac {\kappa}{2}}{\hat a}_+(t)-\sqrt{\kappa}{\hat a}_{\rm in}^{\rm u}(t)\nonumber\\
{\frac{da_-}{dt}}&=&-i\omega_0 {\hat a}_-(t)-{\frac {\kappa}{2}}{\hat a}_-(t)-\sqrt{\kappa}{\hat a}_{\rm in}^{\rm d}(t)\nonumber\\
{\frac{d\sigma_z}{dt}}&=&-\gamma\left(1+\sigma_z(t)\right)\nonumber\\
{\frac{d\sigma^+}{dt}}&=&i\left(E+2\chi {\hat a}^\dagger_+ {\hat a}_+\right)\sigma^+(t)-\left(\gamma_\varphi+{\gamma\over 2}\right) \sigma^+(t)
\end{eqnarray}
where $\left[{\hat a}_{\rm in}^\mu(t), {\hat a}_{\rm in}^{\nu\dagger}(t^\prime)\right]=\delta_{\mu\nu}\delta(t-t^\prime)$ with $\mu, \nu={\rm u,d}$.
Following the standard input-output theory, the output field operator ${\hat a}_{\rm out}(t)$ is related to the input field operator by the following relation: ${\hat a}_{\rm out}(t)=\sqrt{\kappa}{\hat a}(t)+{\hat a}_{\rm in}(t)$. One can notice that the equation of motion for $\sigma_z(t)$ does not couple to the field ${\hat a}_+(t)$ up to the order of $g^2/\Delta$.
We will take the input photon pulse to be a coherent drive at frequency $\omega_r$ and hence its amplitude has a classical and quantum part
\begin{equation}
{\hat a}_{\rm in}^{\mu}(t)=e^{-i\omega_r t}\left[\alpha_{\rm in}(t)+\hat{\xi}^\mu(t)\right]
\end{equation}
where the field operators $\hat{\xi}^{\mu\dagger}(t), \hat{\xi}^\mu(t)$ represent the vacuum noise, which will satisfy the following commutation relations $\left[{\hat\xi}^\mu(t), {\hat\xi}^{\nu\dagger}(t^\prime)\right]=\delta_{\mu\nu}\delta(t-t^\prime)$ with $\mu, \nu={\rm u,d}$\cite{Clerk}. We will assume that $\alpha_{\rm in}(t)$ varies smoothly over $1/\kappa$.
For relatively strong coherent drive, one can neglect the vacuum fluctuation noise delivered to the TSR.

Using the adiabatic approximation, one can obtain the following useful relation between the input and output photon pulses for the input photon pulse with the u-mode
\begin{equation}
\alpha_{\rm out}(t)\cong\frac{i\left(\omega_r-\omega_0-\chi\sigma_z\right)-\kappa/2}{i\left(\omega_r-\omega_0-\chi\sigma_z\right)+\kappa/2}\,\alpha_{\rm in}(t)
\end{equation}
where $\alpha_{\rm in}(t), \alpha_{\rm out}(t)$ represent the slowly varying classical parts of the corresponding photon field operators\cite{Duan}.
By fixing the frequency of the input photon pulse to be $\omega_r=\omega_0-\chi$, one can see that $\alpha_{\rm out}(t)\cong-\alpha_{\rm in}(t)$ for the qubit state with $\sigma_z=-1$, while $\alpha_{\rm out}(t)\cong \alpha_{\rm in}(t)$ for the qubit state with $\sigma_z=+1$. For the input photon pulse with the d-mode, one can show that $\alpha_{\rm out}(t)\cong \alpha_{\rm in}(t)$ independent of the qubit state.
If one prepares the following product state of the superconducting qubit and the flying microwave qubit as an input state described by $\left(\ket{\uparrow}+\ket{\downarrow}\right)/\sqrt{2}\otimes \left(\ket{\rm u}+\ket{\rm d}\right)/\sqrt{2}$, it will be converted to the entangled output state of $\left(\ket{\uparrow {\rm u}}+\ket{\uparrow {\rm u}}-\ket{\downarrow {\rm u}}+\ket{\downarrow {\rm d}}\right)/2$ upon being reflected from our circuit QED system. This represents the two-qubit quantum CPF gate for the hybrid qubit composed of superconducting qubit and flying microwave qubit, which we will call $U_{\rm CPF}$.

\begin{figure}
\epsfxsize=6.5in \epsfysize=4in \epsffile{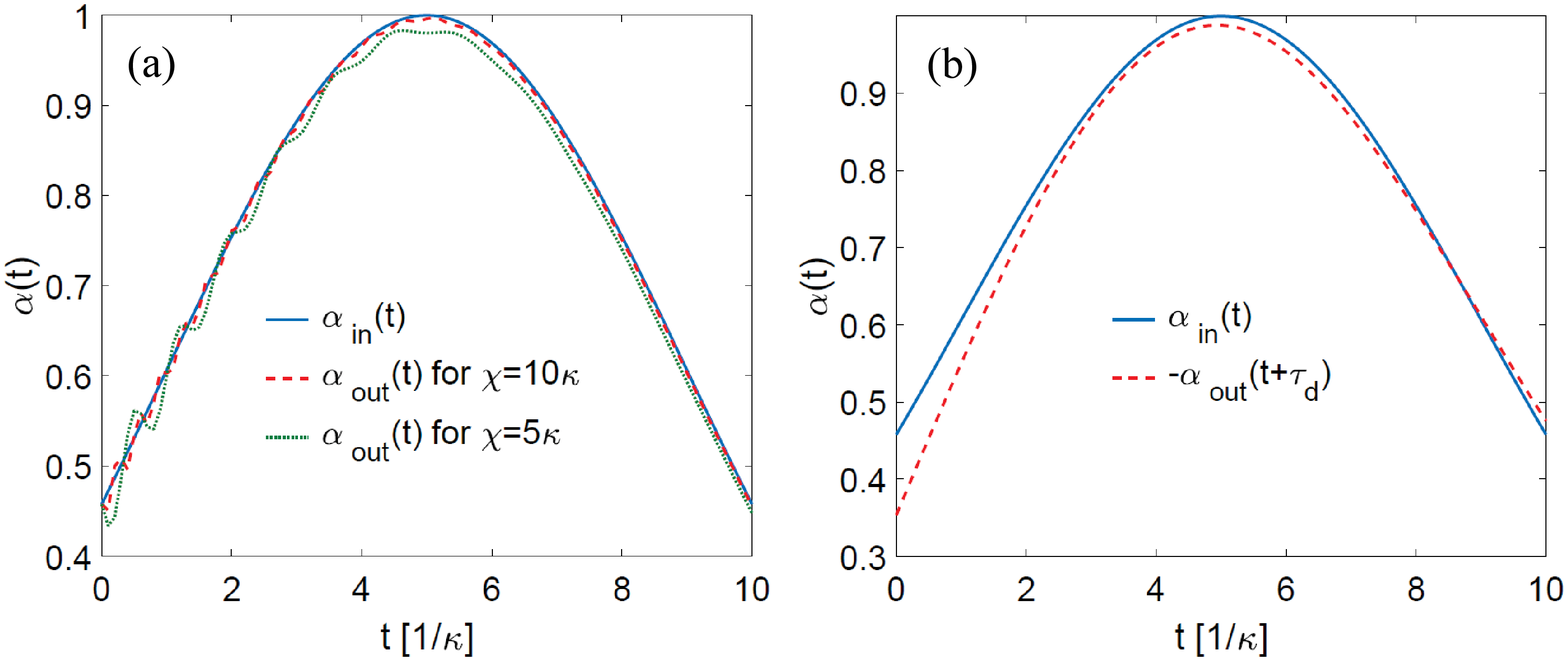}
\caption{Temporal profiles of the input and output photon pulses for the input photon pulse with the u-mode for two different initial qubit states. Temporal profile of the input photon pulse is plotted as a solid curve. (a) For $\sigma_z=+1$ initially, temporal profile of the output photon pulse is plotted as a dashed(dotted) curve for $\chi=10\kappa(5\kappa)$. (b) For $\sigma_z=-1$, temporal profile of the inverted output photon pulse with time delay $-\alpha_{\rm out}(t+\tau_d)$ is plotted as a dashed curve, which is independent of $\chi$. We have taken into account the reflection phase shift of $\pi$ and time delay of $\tau_d\cong 3.5/\kappa$.}
\label{TSR4}
\end{figure}

Using the experimentally available set of parameters based on the standard circuit QED system, we have numerically investigated the effect of finite longitudinal relaxation rate $\gamma$ of the superconducting qubit. For the fixed value of $\omega_r=\omega_0-\chi$, we have chosen the following set of parameters: $\chi=5, 10{\rm MHz}$, $\kappa=1{\rm MHz}$, and $\gamma=0.1{\rm MHz}$\cite{Transmon2,Paik}. We have taken the input photon pulse to be a gaussian pulse centered at $T/2$, which can be written by  $\alpha(t)=e^{-(t-T/2)^2/(2\sigma^2)}/\sqrt{2\pi\sigma^2}$ with $\sigma=4/\kappa$ and $T=1/\gamma$.
We will consider the case that the input photon pulse with the ${\rm u}$-mode is fed into the circuit QED system. The temporal profile of the input photon pulse is plotted as a solid curve in Fig. \ref{TSR4}(a) and \ref{TSR4}(b). For the case of qubit being initially in the excited state, the temporal profile of the output photon pulse is plotted as a dashed(dotted) curve for $\chi=10\kappa(5\kappa)$ in Fig. \ref{TSR4}(a). One can notice that $\alpha_{\rm out}(t)\cong\alpha_{\rm in}(t)$ as expected from the adiabatic approximation.
For the case of qubit being initially in the ground state, the temporal profile of the inverted output photon pulse with time delay $-\alpha_{\rm ou}(t+\tau_d)$ is plotted as a dashed curve in Fig. \ref{TSR4}(b), which is independent of $\chi$. We have taken into account the reflection phase shift of $\pi$ and time delay of $\tau_d\cong 3.5/\kappa$. It demonstrates that $\alpha_{\rm out}(t+\tau_d)\cong -\alpha_{\rm in}(t)$. In both cases, the deviations in temporal profiles of the output photon pulses from the input photon pulse are relatively small.
In the LOQC scheme using an optical cavity QED, it has been demonstrated that the two-photon quantum CPF gate can be implemented by applying a series of operations composed of the CPF gate for the hybrid qubit applied to two individual input photon pulses and the single-qubit rotation $U_{\rm Q}$:
$U_{\rm TQCPF}=U_{\rm CPF}^{\rm 1st}\,U_{\rm Q}(-\pi/2)\,U_{\rm CPF}^{\rm 2nd}\,U_{\rm Q}(\pi/2)\,U_{\rm CPF}^{\rm 1st}$\cite{Duan,Hacker}.

To summarize, we have introduced the new circuit QED system with a triple-leg stripline resonator, which can support two-fold degenerate intra-cavity microwave photon modes.
Placing a qubit at one of the legs of the TSR makes one photon mode couple to the qubit, while the other mode remains uncoupled.
We have utilized this interesting property of two intra-cavity microwave photons to realize the CPF gate for the hybrid two-qubit state composed of superconducting qubit and flying microwave qubit using the linear optical quantum computing scheme. Through the extensive numerical simulations based on the experimentally available set of parameters, we have theoretically demonstrated that the CPF gate for the hybrid qubit can be reliably implemented. One interesting aspect of the TSR system is that one can build up a photonic honeycomb lattice using the TSR as a basic building block, since the TSR has triple-legs.

K.~M. wishes to acknowledge the financial support by Basic Science Research Program through the National Research Foundation of Korea (NRF) funded by the Ministry of Education, Science and Technology (NRF-2016R1D1A1B01013756).

\end{document}